\documentclass[12pt, a4paper, lineno]{article}
\usepackage{lineno}
\usepackage{amsmath, amssymb, theorem}
\usepackage{setspace}
\usepackage{rotate, color}
\usepackage{graphicx}
\usepackage{chicago}
\newcommand{\be}{\begin{equation}}
\newcommand{\ee}{\end{equation}}  
\newcommand{\bean}{\begin{eqnarray*}} 
\newcommand{\eean}{\end{eqnarray*}} 

\newtheorem{theorem}{Theorem}
\newtheorem{remark}{Remark}

\newtheorem{lemma}[theorem]{Lemma}

\theoremstyle{definition}

\theoremstyle{remark}
\newtheorem{rem}[remark]{Remark}
\theoremstyle{proof}

\newcommand{\X}{\mathbf{X}}
\newcommand{\Y}{\mathbf{Y}}
\newcommand{\N}{\mathbf{N}}
\newcommand{\V}{\mathbf{V}}
\newcommand{\T}{\tau}
\newcommand{\inD}{\rightarrow_{\mathcal L}}
\newcommand{\opt}{\mbox{\tiny opt}}
\newcommand{\purple}[1]{\textcolor{black}{#1}}
\definecolor{purple}{rgb}{0.7,0.0,0.8}
\newcommand{\red}[1]{\textcolor{black}{#1}}
\newcommand{\blue}[1]{\textcolor{black}{#1}}

\begin{document}



\title{Sequential estimation for covariate-adjusted response-adaptive designs \\
   \vspace{5mm}
\large{Running headline: Sequential estimation for CARA designs}}

\author{Yuan-chin Ivan Chang\\Institute of Statistical Science, Academia Sinica, Taipei
11529, Taiwan \\
Eunsik Park\\Department of Statistics, Chonnam National University,\\
Gwangju 500-757, Korea }


\maketitle

\newpage
\begin{abstract}
In clinical trials, a covariate-adjusted response-adaptive (CARA) design allows a subject newly entering a trial a better chance of being allocated to a superior treatment regimen based on cumulative information from previous subjects, and adjusts the allocation according to individual covariate information.
\blue{Since this design allocates subjects sequentially, it is natural to apply a sequential method for estimating the treatment effect \red{in order to make the data analysis more efficient}.}
In this paper, we study the sequential estimation of treatment effect for a general CARA design. A stopping criterion is proposed such that the estimates satisfy a prescribed precision when the sampling is stopped. The properties of estimates \blue{and stopping time} are obtained under the proposed stopping rule. In addition, we show that the asymptotic properties of the allocation function\red{, under the proposed stopping rule,} are the same as those obtained in the non-sequential/fixed sample size counterpart.
%
We then illustrate the performance of the proposed procedure with some simulation results using logistic models.  The properties, such as the coverage probability of treatment effect, correct allocation proportion and average sample size, for diverse combinations of initial sample sizes and tuning parameters in the utility function are discussed.

\vspace{0.2 in}
\end{abstract}

\renewcommand{\thesection}{\arabic{section}.}
\setcounter{section}{0}
\section{Introduction}
\renewcommand{\thesection}{\arabic{section}}

From an ethical viewpoint, it is desirable to minimize the number of subjects allocated to inferior treatments in the course of a clinical trial 
 without jeopardizing the generation of useful and meaningful statistical inferences. The response adaptive (RA) design in clinical trials (\citeNP{ZW95} and \citeNP{HU-rosen06}) is dedicated to this purpose.
The advantage of an RA design is that the information collected from subjects previously entering the trial can be used to adjust the allocation probability so that a newly entering subject can have a better chance of being allocated to a superior treatment. Because of the sequential characteristic in this process, sequential statistical methods should be used \red{in order to efficiently analyze} these kinds of data sets. Since data collected in this manner are no longer independent, sequential methods that rely on assumption of independent observations are not valid.
Moreover, due to innovation in genomic technologies and the nature of developing targeted drugs \shortcite{SM05}, it is natural to incorporate the information available on individual covariates that have a strong influence on responses to a model, since they may be associated with the efficacy of treatments.
{Hence, the existence of an interaction between treatment and covariate becomes a reasonable presumption as \blue{far as}, for example, a targeted drug is concerned.}

A situation where there is an interaction between covariates and treatments is illustrated {in Figure 1. In this figure,} {a logistic model is used to describe the relation between responses to treatments and covariates, where the covariates are generated} from two normal distributions with a mean-shift denoting two sub-populations.
Traditionally, we use an RA design by assuming there is no treatment-covariate interaction effect\red{;} that is, the slopes of treatments effects are assumed to be equal. However, when a treatment-covariate interaction exists, as in Figure 1, this assumption is not valid, and the lines A and B in this Figure will not be parallel.
This implies that a method that uses RA design will make incorrect treatment allocation, {when such a non-ignorable interaction exists.
In this situation, it is reasonable to assume that a CARA design should perform better than an RA design in terms of correct allocation proportions. However, up until now, little work has been done on CARA designs. Since Figure 1 is for illustration purposes, it depicts an extreme example of two treatments with opposite slopes. However, as long as the slopes of the treatment effects are not equal, the two treatments make a lot of difference for subjects with covariates located far from the intersection of the lines of the treatment effects. Thus, as long as a targeted drug or other adaptive treatment strategy is being used, this situation should not be ignored.  In addition to the ethical considerations, this is a further good reason for considering a CARA design.}
{Further discussion about the properties of RA and CARA designs can be found in, \citeN{BBB07}, \citeN{Bandyopadhyay09}, \citeN{HU-rosen06} and so on.}

Although the sequential characteristics of RA and CARA designs are clear, {and the sequential sampling method, which allows the sample size to be determined based on the observed information, is known to be an adequate choice for making efficient and valid statistical inference, most discussions in the literature to date have been limited to the asymptotic properties of different designs. Even when the idea of a stopping rule has been adopted, there has still been very little discussion of estimation under those stopping criteria.  \citeN{zhang-Hu09}, and \citeN{Bandyopadhyay09} are two typical examples. In these two studies, only large scale simulation studies were conducted to compare the properties of {their} designs and to provide information regarding suitable sample sizes for their designs. In another example, \citeN{Moler06} treated the allocation ruled by an urn model as a Robbins-Monro scheme, but the property of the stopping rule was still ignored. In addition, \citeN{Thanll-wathen05} compared the CARA design to the balanced randomization design, however, the same stopping rule based on the balanced randomized design was applied to both designs, which is inappropriate as indicated in their paper.}

{\purple{As mentioned above and also in \citeN{HU-rosen06}, the sequential method is a natural choice \blue{for} a CARA design based clinical trial; however,} it is rare to find literature regarding the application of  stopping rules for the sequential estimation procedure based on CARA designs, and the effective sample size for a clinical trial with adaptive design. \purple{The difficulties} are mostly due to the adaptive nature of CARA designs, which make the classical approach, based on the assumption of independent observation, less useful. Besides the adaptive design, the adjustment of the allocation probability based on subject's covariate information makes the procedure even more complicated. Hence, the asymptotic properties of estimates under randomly stopped CARA experiments, derived in our paper, are not trivial and cannot follow from their non-random sample size counterpart.}

{In this paper, a sequential procedure is proposed for estimating treatment effect under a general CARA design. \blue{Our goal is to estimate the treatment effects}, \red{with the minimum sample size,} such that the estimates satisfy a prescribed precision, and subjects can be allocated to the superior treatment without interfering with the the quality and efficiency of estimation of treatment effects. The asymptotic properties of sequential estimates are obtained under this general CARA design.  In addition, we also show that the allocation rule\red{, under the proposed stopping criterion,} maintains the same asymptotic properties as those obtained in its non-sequential counterpart. In our numerical study, for illustration purposes, we adopt the method of \shortciteN{BBB07} and use a utility function} \blue{to balance the ethical consideration and the efficiency of the estimate for treatment allocation. We, then, modify the utility function to vary the tuning parameters} {sequentially depending on the precision of the estimate at every allocation stage such that subjects are allocated to a ``more adequate'' treatment.}

The rest of this paper is organized as follows: A sequential estimation procedure for treatment effect is proposed in Section 2. Simulation results are applied to logistic models using a modified allocation rule \red{(\citeNP{BBB07})}  in Section 3. We, then, conclude with discussion in Section 4. Proofs of theorems are given in the Appendix.



\renewcommand{\thesection}{\arabic{section}.}
\setcounter{section}{1}
\section{Sequential Estimation of Treatment Effect}
\renewcommand{\thesection}{\arabic{section}}\label{sec: SETE}

Let $N_{m, k}$ be the number of subjects assigned to treatment $k$ during the first $m$ assignments and ${\N}_m = (N_{m,1}, \ldots, N_{m,K})$.  Suppose that $\{ Y_{m,k}, m=1, 2, \ldots, k=1, \ldots, K\}$ denotes responses of the $m$-th subject to the $k$-th treatment and ${\Y}_m= (Y_{m,1}, \ldots, Y_{m, K})$. Let ${\boldsymbol\xi}_m$ be the covariates of the $m$-th subject.
{Suppose that $\X_1, \X_2, \ldots $ is the sequence of random treatment assignments, and
$\X_m = (X_{m,1}, \ldots, X_{m,K})$, $X_{m,k} \in \{0, 1\}$,
denotes assignment of treatment $k$ to the $m$-th subject.
Then $X_{m,k}=1$ {for some $k$} and $\sum_{k=1}^K X_{m,k}=1$.
That is, each subject is allocated to one treatment only.
Hence, it follows that the response of subject $m$ to the treatment $k$, $Y_{m,k}$, is observed only if $X_{m,k}=1$.
(Note that this implies that ${\N}_m=\sum_{i=1}^m X_i$.)}

Define $\mathcal{X}_m =\sigma(\X_1, \ldots, \X_m)$, $\mathcal{Y}_m = \sigma({\Y}_1, \ldots, {\Y}_m)$, and $\mathcal{Z}_m=\sigma({\boldsymbol\xi}_1, \ldots, {\boldsymbol\xi}_m)$, $\boldsymbol\xi_i \in \mathbb{R}^p$, be the corresponding $\sigma$-fields.
Let $\mathcal{F}_m=\sigma(\mathcal{X}_m, \mathcal{Y}_m, \mathcal{Z}_m)$, then a general CARA design is defined as
\[
    {\boldsymbol\psi}_m=E[\X_m| \mathcal{F}_{m-1}, \boldsymbol\xi_m]=E[\X_m| \mathcal{X}_{m-1}, \mathcal{Y}_{m-1}, \mathcal{Z}_{m}.]
\]
Suppose that for each $m \geq 1$, the responses and covariate vector satisfy
\begin{align}\label{model-1}{
  E[Y_{m,k}|\boldsymbol\xi]=\mu_k(\theta_k, \boldsymbol\xi),
  }
\end{align}
where  \purple{$\mu_k(\cdot , \cdot)$ are known functions, $V_k$  denotes the covariance matrix based on Equation (\ref{model-1}) and $\theta_k\in R^p$ for $k=1, \ldots, K$.}
The asymptotic properties of the estimate of {$\boldsymbol\theta=(\theta_1, \ldots, \theta_K)$} and allocation function under such a general CARA design has been discussed in \shortciteN{ZFCC07}. The estimation of $\theta$ is the primary goal in a clinical trial.  Thus, it will be beneficial if treatment effects can be estimated with a certain accuracy using a minimum required sample size whilst simultaneously still retaining the good allocation properties. Since, in a CARA design, the design at the current stage depends on the past history, sequential analysis is the statistical tool of choie. \red{Here} a sequential estimation procedure is proposed for constructing a confidence set for $\theta$ with a prescribed accuracy, and we show that the asymptotic properties of allocation function remain the same as their non-sequential counterparts under such a sequential sampling strategy.

Suppose no prior information about the effects of treatments is available. In order to estimate the treatment effects,  at the beginning, we need to assign $m_0 (>0)$ subjects to each treatment using restricted randomization. Hence, when we allocate the $m$-th subject ($m > Km_0$), there are already $m-1$ observations, $\{(\X_1, \Y_1, \boldsymbol\xi_1), \ldots, (\X_{m-1}, \Y_{m-1}, \boldsymbol\xi_{m-1})\}$, collected. Thus, we assign the $m$-th subject to the treatment $k$ with probability
\[
    \psi_k=P(X_{m,k}=1 | \mathcal{F}_{m-1}, \boldsymbol\xi_m)=\pi_k(\hat\theta_{m-1}, \boldsymbol\xi_m),
\]
where $\hat\theta_{m-1}$ is the \blue{maximum quasi-likelihood} estimate of $\theta$ based on the previous $m-1$ observations \blue{and $\pi_k(\cdot, \cdot)$ is the true allocation probability for treatment $k$ and the given covariate}.
Assume further that $\mu_k(\theta_k, \boldsymbol\xi_m) = \mu_k(\boldsymbol\xi_m'\theta_k)$ for each $m \geq 1$.
Hence, it follows from Equation (1) and $V$, that the method of generalized linear models (quasi-likelihood) can be applied \cite{McCullagh89}. Assume that $\theta_k \in  \Theta_k \subseteq R^p$ is bounded for $k=1, \ldots, K$, and let the parameter space $\Theta=\prod_{k=1}^K\Theta_k$.

Under the above assumptions (see also Condition A of \shortciteN{ZFCC07}, Theorem 2.1), it is proved that as $\min(N_{m,k}, k=1, \ldots, K)$ goes to infinity,
\[
   \red{ \sqrt n ( {\hat {\boldsymbol\theta}} -\boldsymbol \theta) \blue{\inD} N(0, {\V}),}
\]
where \red{$\V=\hbox{diag}\{V_1, \ldots, V_K \}$.}
{Based on the asymptotic normality of $\boldsymbol{\hat\theta}$, the sequential method is employed for estimating the confidence set of $\boldsymbol\theta=(\theta_1, \ldots, \theta_K)$.}  Define
\[
 \blue{R = \{\boldsymbol\theta \in \Theta : n (\hat{\boldsymbol\theta} - \boldsymbol\theta)' \V^{-1} (\hat{\boldsymbol\theta} -\boldsymbol\theta) \leq C^2_\alpha
 \},}
\]
where $C^2_\alpha$ is the constant such that $P(\chi^2 (p\cdot K )
\geq C^2_\alpha) \leq \alpha$.
The asymptotic normality of $\hat {\boldsymbol\theta}$ implies that $P(\boldsymbol\theta \in R) \approx 1-\alpha$ as the sample size becomes large.

Although large sample results  guarantee the performance of estimates and some asymptotic properties of CARA designs, we want to know just how large a sample size is needed to
guarantee a satisfactory performance in a practical sense.
Moreover, no matter how high the coverage probability
is, the confidence set becomes less useful if {the size of the confidence set} becomes too
large.  Now, suppose we further require that the maximum axis of $R$ is no larger than $2 \delta$ for some $\delta > 0$, then the minimum sample size to achieve this goal is
\[ n \Lambda_{\min} ({\V}^{-1}) \geq \frac{C^2_\alpha}{\delta^2}. \]
Equivalently, the above inequality can be re-written as
\begin{align}
  \label{optimal-sample-size}
  n \geq \frac{C^2_\alpha \Lambda_{\max}(\V)}{\delta^2},
\end{align}
where notations $\Lambda_{\max}(A)$ and $\Lambda_{\min}(A)$ denote the maximum and minimum eigenvalues of matrix $A$, respectively.
{Let $R_\delta$ denote the corresponding confidence ellipsoid for given $\delta$. So, once $\delta >0$ is specified, the maximum axis of confidence ellipsoid $R_\delta$ is no greater than $2\delta$. The constant $\delta$ here is used as a measure of precision of the confidence ellipsoid $R_\delta$. Please refer to \citeN{siegmund85}, \citeN{albert66} and \citeN{GhoshSen91} for other measures of confidence sets.}

If $\V$ is known, then the optimal sample size required to
construct a confidence ellipsoid $R_\delta$ with the required
maximum axis no greater than $2 \delta$ is
\[
    n_{opt} = \mbox{ first } n \mbox{ such that } n \geq \frac{C^2_\alpha \Lambda_{\max}(\V)}{\delta^2}.
\]
Since the variance matrix $\V$ is usually unknown, the above
optimal sample size is not available. Replacing the unknown $\V$ in Equation (\ref{optimal-sample-size}) with its {consistent} estimate
$\boldsymbol{\hat \theta}$ (to be defined later), a stopping
rule to construct such a fixed size confidence ellipsoid is suggested:
\begin{align}
  \label{stopping-rule-original-parameter}
  \T_\delta &= \mbox{ first } n \mbox{ such that } n \geq \frac{C^2_\alpha \Lambda_{\max}({ \hat \V})}{\delta^2}\notag \\
           &= \inf\{ n \geq n_0 : n \geq \frac{C^2_\alpha \Lambda_{\max}({ \hat \V} )}{\delta^2}\},
\end{align}
where $n_0 \geq Km_0$ is the minimum initial sample size and $m_0$
is the initial sample size for each treatment. Similarly, we then
define
\[
 \hat R_\delta= \{\boldsymbol\theta \in \Theta : n (\hat{\boldsymbol\theta} - \boldsymbol\theta)' {\hat \V}^{-1}
 (\hat{\boldsymbol\theta} -\boldsymbol\theta) \leq C^2_\alpha \}.
\]
It follows from the strong consistency of $\hat {\boldsymbol\theta}$, if ${\hat \V}$ is also a strongly consistent estimate of $\V$, then
$\lim_{n\rightarrow \infty}P(\boldsymbol\theta \in \hat R_\delta) = 1 -\alpha$.
That is, $\hat R_\delta$ is a confidence ellipsoid of $\boldsymbol\theta$ with
coverage probability $1-\alpha$, asymptotically.

It follows from the definition of $\T_\delta$ that, when the {sequential} sampling stops, the confidence ellipsoid will have its maximum axis no greater than $2 \delta$. However, it is also known that there is no guarantee that $\hat {\boldsymbol\theta}$ will have the same asymptotic distribution if we replace the fixed sample size with a random sample size $\T_\delta$.  Although the sequential estimation procedure provides a way to control the size of the confidence set by utilizing a stopping rule, it is interesting to know whether \red{the} asymptotic properties in \shortciteN{ZFCC07} are still adhered to under such a randomly stopped criterion.

{
Suppose that allocation function $\boldsymbol{\pi}(\cdot, \cdot) =\left( \pi_1(\cdot, \cdot), \ldots, \pi_K(\cdot, \cdot) \right)$ and satisfies the following conditions:
\begin{itemize}
  \item[] (C1) $\sum_{k=1}^K \pi_k=1$ and $0< \nu_k=E_{\boldsymbol\xi}[\pi_k(\boldsymbol\theta, \boldsymbol\xi)] <1$, $k=1, \ldots, K$.
  \item[] (C2) For fixed $\boldsymbol\xi$, $\pi_k(\boldsymbol\theta, \boldsymbol\xi) >0$ is a continuous function of $\theta$ and is differentiable with respect to $\theta$ such that $\nu_k(\boldsymbol{\tilde \theta})=\nu_k(\boldsymbol\theta) + (\boldsymbol{\tilde\theta} -\boldsymbol\theta)(\partial \nu_k/\partial\boldsymbol{\tilde\theta})' +o(\|\boldsymbol{\tilde\theta} - \boldsymbol\theta \|^{1+\zeta}) $ for some $\zeta >0$.
\end{itemize}
\blue{The condition $\pi_k > 0$ for each $k=1, \ldots, K$ on the allocation function guarantees that subjects will be allocated to individual treatments, eventually. Thus, this condition also affirms that with probability one the design matrix is non-singular, and the $\Lambda_{\min}(\V^{-1}) > 0$, asymptotically.}
\blue{Under these conditions,} in Theorem \ref{thm:seq-main}, we show that the sequential procedure with the stopping rule defined in (\ref{stopping-rule-original-parameter}) can guarantee that the size of the maximum axis of confidence ellipsoid is no greater than the pre-specified length, while maintaining the required coverage probability.
\blue{In addition to} classical asymptotic properties of sequential confidence set estimation, the asymptotic properties of the allocation function under sequential sampling that is based on the CARA design are also proved \blue{in Theorem \ref{thm:seq-main}}.
}
\begin{theorem}
  \label{thm:seq-main}
  {Under some regularity conditions on the link function $\mu_k$ and Conditions (C1) and (C2) for the allocation function $\nu_k$, for each $k$, if $\sup_m \| \boldsymbol\xi_m \| < \infty$, then} the proposed sequential estimation with the stopping rule defined in (\ref{stopping-rule-original-parameter}) has the following properties:
  \begin{itemize}
    \item[](i)
        $P(\T_\delta < \infty) =1$ and $\lim_{\delta\rightarrow 0}\T_\delta/n_{\opt} = 1$
    almost surely.
  \end{itemize}
  When the sampling stops, the estimate of $\theta$ satisfies that
  \begin{itemize}
    \item[](ii)
        $\boldsymbol{\hat \theta}_{\T_\delta} \rightarrow \boldsymbol\theta$ almost surely as $\delta \rightarrow 0$,
        $ \sqrt {\T_\delta} ({\boldsymbol{\hat \theta}_{\T_\delta}} - \boldsymbol\theta)
        \inD N(0, {V})$,
  and
        $\lim_{\delta \rightarrow 0}P(\blue{\boldsymbol\theta} \in R_\delta) = 1-\alpha$.
  \end{itemize}
  Then, in addition, the average of the stopping rule satisfies that
  \begin{itemize}
    \item [](iii) $\lim_{\delta \rightarrow 0} E\left[\frac{\T_\delta}{n_{\opt}}\right] =1$.
  \end{itemize}
  Moreover, for a given allocation function, 
  it is shown that
  \begin{itemize}
    \item[](iv) $\lim_{\delta \rightarrow 0} \frac{\N_{\T_\delta}}{\T_\delta}=\boldsymbol\nu$ almost {surely,}
    \item[] (v) $\frac{N_{\T_\delta,k|\blue{\boldsymbol\xi}}}{N_{\T_\delta|\boldsymbol\xi}}\rightarrow
    \pi_k(\boldsymbol\theta,\boldsymbol\xi) \mbox{ a.s. \blue{as $\delta \rightarrow 0$}, } k=1, \ldots, K$, and
    \item[] {(vi) $\sqrt{\T_\delta} (\N_{\T_\delta}/{\T_\delta} - \boldsymbol\nu) \inD N(\mathbf{0}, \mathbf{\Sigma})$,}
  \end{itemize}
  where ${N_{\T_\delta,k|\blue{\boldsymbol\xi}}}$ is the number of subjects assigned to treatment $k$ with covariate $\boldsymbol\xi$ up to $\T_\delta$th subject and ${N_{\T_\delta|\blue{\boldsymbol\xi}}}$ is the total number of subjects with covariate $\blue{\boldsymbol\xi}$  up to $\T_\delta$th subject. Here \purple{$\boldsymbol\nu =(\nu_1, \cdots, \nu_K)'$} and $\pi_k$, $k=1, \ldots, K$, depend on the allocation function,
  {and $\boldsymbol\Sigma=\Sigma_1 + 2\Sigma_2$ where $\Sigma_1 = diag\{\boldsymbol\nu\} - \boldsymbol{\nu{'}\nu}$ and $\Sigma_2=\sum_{k=1}^K  (\frac{\partial \boldsymbol\nu}{\partial \theta_k}) V_k (\frac{\partial \boldsymbol\nu}{\partial \theta_k})'$.}

\end{theorem}
Theorem \ref{thm:seq-main} (i) states that the sequential sampling
will stop eventually, and (ii) and (iii) are named asymptotic
consistency and efficiency of a sequential confidence estimation
procedure by \shortciteN{chow-robbins65}. Theorem
\ref{thm:seq-main} (iii) means that the average ratio of the
sequential sample size to optimal sample size converges to 1. This
means the proposed sequential sampling is efficient in terms of
sample size used for constructing a fixed size confidence
ellipsoid of the parameters of interest.

{Theorem \ref{thm:seq-main} (iv) to (vi) provides the asymptotic properties of the allocation rule under the sequential estimation procedure. In particular, Theorem \ref{thm:seq-main} (iv) states that eventually the allocation proportion converges to the allocation expectation $\nu$, and Theorem \ref{thm:seq-main} (v) states that for the given covariate $\boldsymbol\xi$, the proportion of allocation converges to the ``true'' (unknown) allocation probability with probability one as $\delta$ goes to zero.  That is, if the conditions in Theorem \ref{thm:seq-main} are satisfied, then under the proposed sequential sampling method,  the allocation rule maintains the same asymptotic properties as those in its non-sequential sampling counterpart.
In our simulation study, we have demonstrated our procedure using the allocation rule proposed in \shortciteN{BBB07}.  Please refer to \shortciteN{ZFCC07} for different allocation functions/designs under this general framework. }

%

\begin{rem}
  Note that the proof of the properties of the sequential procedure is not trivial, and cannot follow directly from the results of the estimates based on the non-random sample size case due to the application of the stopping rule. This can be seen from a simple example in \citeN{ChowTeicher88} (Chapter 4, Example 1, page 90). {Since} our proof of Theorem \ref{thm:seq-main} is based on the last time approach of \citeN{chang99}, some conditions on the parameter space $\Theta$ can be relaxed.  Details are given in the Appendix.
\end{rem}

\subsection{Subset of parameters}
\label{s:subset} Sometimes, we are only interested in contrasts of
parameters. For example, instead of estimating individual
treatment effects, we may want to estimate {differences} between
treatment effects in a clinical trial with multiple {treatments. For} this purpose, let $H$
be a $p\times h$ matrix that specifies the contrasts with $0 <
\hbox{Rank}(H)= h \leq p$. Let $\boldsymbol\gamma = H'\boldsymbol\theta$, then the
asymptotic properties of $\boldsymbol{\hat\theta}$ imply that as $n
\rightarrow \infty$
\[
    \sqrt n (\hat {\boldsymbol\gamma}_n - \boldsymbol\gamma) \inD N(0, \V_\gamma),
\]
where $V_\gamma = H'\V H$. Let $\hat \V_\gamma = H'\hat V H$.  Then $\hat \V_\gamma$ is a strongly consistent estimate of $\V_\gamma$. Therefore, it follows that $n(\boldsymbol{\hat\gamma} -  \boldsymbol\gamma)' \hat \V^{-1}_\gamma (\hat {\boldsymbol\gamma} -  \boldsymbol\gamma)$ is asymptotically distributed with $\chi^2 (h)$. Let
\begin{align}\label{eq:con-subset}
 R_{\delta_\gamma} = \{\boldsymbol\gamma \in R^h : n(\hat{\boldsymbol\gamma} -  \boldsymbol\gamma)'
 \V^{-1}_\gamma (\hat{\boldsymbol\gamma} -  \boldsymbol\gamma) \leq C^2_{\alpha, \gamma}\}
\end{align}
Similarly, we can also construct a confidence ellipsoid of $\boldsymbol\gamma$ with the length of its maximum axis no greater than $2\delta_\gamma$. Then the optimal sample size and its corresponding stopping time are
 \begin{align}\label{eq:optimal-subset}
    n_{\gamma,opt} = \mbox{ first } n \mbox{ such that } n \geq
  \frac{C^2_{\alpha, \gamma} \Lambda_{\max}(\V_\gamma)}{\delta_\gamma^2}
 \end{align}
  and
 \begin{align}\label{eq:stop-subset}
    \T_{\delta_\gamma}
           = \inf\{ n \geq n_0 : n \geq \frac{C^2_{\alpha, \gamma}
           \Lambda_{\max}({ \hat \V_\gamma} )}{\delta_\gamma^2}\}.
 \end{align}
{By simple matrix algebra, we have a parallel theorem to Theorem \ref{thm:seq-main} for
contrasts of parameters.}
\begin{theorem}\label{thm:subset} {Let $H$ be a $p\times h$ matrix with Rank$(H)=h \leq p$,
  and $\boldsymbol\gamma = H'\boldsymbol\theta$.
  Then under conditions similar to Theorem \ref{thm:seq-main}, $\hat {\boldsymbol\gamma}$ is a strongly consistent estimate of $\boldsymbol\gamma$ and asymptotically normally distributed with covariance matrix $\V_\gamma=H'\V H$.  Moreover, the sequential procedure with the stopping rule defined in (\ref{eq:stop-subset}) has the following asymptotical properties:
    \begin{itemize}
    \item[] (i) $P(\T_{\delta_\gamma} < \infty)=1$ and $ \lim_{\delta_\gamma \rightarrow
    0}\T_{\delta_\gamma}/n_{\gamma, \opt} =1$ almost surely.
    \item[] (ii)
 $ \boldsymbol{\hat\gamma}_{\T_{\delta_\gamma}} \rightarrow \boldsymbol\gamma$ almost surely as ${\delta_\gamma} \rightarrow 0$,
    $ \sqrt {\T_{\delta_\gamma}} ({ \boldsymbol{\hat\gamma}_{\T_{\delta_\gamma}}} - \boldsymbol\gamma) \inD N(0, {\V_{\gamma}})$, and
    $\lim_{\delta_\gamma \rightarrow
    0}P(\boldsymbol\gamma \in R_{\delta_\gamma})= 1-\alpha$.
    \item[] (iii) $\lim_{\delta_\gamma \rightarrow
    0} E\left[\frac{\T_{\delta_\gamma}}{n_{\gamma, \opt}}\right]=1$,
  \end{itemize}
  where $R_{\delta_\gamma}$ and $n_{\gamma,opt}$ are defined in (\ref{eq:con-subset}) and (\ref{eq:optimal-subset}), respectively.
  }
\end{theorem}

The main difference between the new stopping rule defined in Equation(\ref{eq:stop-subset}) and the previous one is the variance of $\hat{\boldsymbol\gamma}$, and this difference in $\T_{\delta_\gamma}$ does not affect the allocation rule. Therefore, the asymptotic properties of the allocation rule in Theorem \ref{thm:subset} follow from the same arguments as in the proof of Theorem \ref{thm:seq-main}. In fact, the asymptotic properties of the allocation rule remain the same under this stopping rule, and are not re-stated here.  That is, this sequential estimation procedure allows us to compare treatment effects using a contrast estimation method under a CARA design without disturbing the asymptotic properties of the allocation function, which is a useful feature in practice.
\begin{rem}
    Note that the asymptotic properties of the allocation function in Theorem \ref{thm:subset} will remain the same as those in Theorem \ref{thm:seq-main} when $\delta_\gamma$ becomes small. However, intuitively, the sequential sample sizes should converge at different rates, {depending on} the contrasts. {This property is usually reflected} in the second order term of the stopping time and is not shown in {Theorem \ref{thm:subset}}.
\end{rem}

\renewcommand{\thesection}{\arabic{section}.}
\setcounter{section}{2}
\section{Numerical Study}\label{sec: NS}
\renewcommand{\thesection}{\arabic{section}}

The purpose of the numerical study is to look at the performance of the estimate of the treatment effect and the allocation of subjects.
In order to apply the sequential confidence estimation procedure proposed in Section \blue{2} 
for $K$ treatments, and treatment allocation procedures in Section \blue{\ref{sec: TAR}}, 
for illustration purposes, we consider a binary response case in this study using the logistic model.

\subsection{Treatment Allocation Rule}\label{sec: TAR}
{In order to skew the treatment allocation proportion so that the better treatment is allocated more often, \shortciteN{BBB07} suggests using an utility function below. For $K$ treatments, their utility function is defined as}
\begin{align}\label{utility}
  U(p) = \log |\hat I_{n+1}| - \eta \left\{ \sum_{k=1}^K p_k
  \log\left( \frac{p_k}{\pi_k(\boldsymbol {\hat\theta}, \boldsymbol\xi)} \right) \right\},
\end{align}
where $\pi_k( \boldsymbol{\hat\theta}, \boldsymbol\xi)$ is the estimate of $\pi_k(\boldsymbol\theta, \boldsymbol\xi)$ denoting the estimate of the allocation probability for treatment $k$ up to current stage $n$.
For a given $\boldsymbol\xi$ and the current estimate of $\boldsymbol\theta$, the optimal allocation rule is to find the vector of probabilities $\boldsymbol p=(p_1, \ldots, p_K)$ that maximize the utility function above. That is, the design at the $(n+1)$th stage is to allocate the $(n+1)$th subject to the treatment that maximizes the utility function.

In the utility function, the first term is in log $n$ scale, which is a log determinant of the information matrix. If $\eta=0$, then the new subject is selected to maximize the Fisher information matrix, which is referred to as the piecewise D-optimal design as mentioned in \shortciteN{BBB07}.
On the other hand, if $\eta$ goes to $\infty$, then the optimal value of $p$ is to maximize the relative entropy function, the second term of (\ref{utility}), which was also raised in \shortciteN{Band-Biswas2001}.
Hence, the parameter $\eta$ can be used to adjust the ethical and efficiency balance.
Here we use \blue{a} utility function to balance the needs for estimation precision of treatment effects and the ethical consideration. It leads to the (locally) D-optimal design.

At the beginning of a study, when estimates of treatment effects are not reliable, we can improve the precision of the estimation of treatment effects when allocating patients via a utility function. Since the estimate of treatment effects becomes stable as the sample size becomes large, it is reasonable to move the weight gradually toward the ethical part at the later stage of the study. If there is sufficient information on treatment effects, we tend to allocate more patients to the better treatment. That is, unlike the two-stage design in \shortciteN{BBB07}, we now have more flexibility to alter the parameters of the utility function as sampling goes on such that the needs for estimating treatment effects and the ethical consideration can be fulfilled and balanced.

The second term in the utility function involves $\pi_k( \boldsymbol{\hat\theta}, \boldsymbol\xi)$. Modifying the utility function by \shortciteN{BBB07}, \blue{$\pi_k(\boldsymbol{\hat\theta}, \boldsymbol\xi)$} can be defined as follows with $K=2$ for illustration purposes.
\[ {
    \pi_1( \boldsymbol{\hat\theta}, \boldsymbol\xi)=
    J\left(
    \frac{\boldsymbol{\xi'\hat\theta}_1 - \boldsymbol{\xi'\hat\theta_2}}{T_n}
    \right)
    ~~\rm{and}~~ \pi_2(\boldsymbol{\hat \theta}, \boldsymbol\xi)=1-\pi_1(\boldsymbol{\hat \theta}, \boldsymbol\xi),
    }
\]
where $J(t)$ can be any symmetric function.
$\pi_k(\boldsymbol{\hat \theta}, \boldsymbol\xi)$ can vary sequentially through $T_n$ at each allocation.
Both $T_n$ and $\eta$ can serve as tuning parameters between efficiency and ethics and be random depending on the estimate precision, which can be a function of standard deviation of the treatment effect estimate based on cumulative observations up to $n$th subject. Please note that $T_n$ and $\eta$ are also tuned by a new covariate $\xi$ of the $(n+1)$th subject.
\blue{Through numerical studies, \purple{\shortciteN{BBB07} provides tables with estimates of allocation proportions for several $\eta$s and given $T_n$ for two stage CARA designs}. In Section \ref{sec:simu}, we \purple{present} numerical results with \purple{some suggestions for tuning both parameters of $T_n$ and $\eta$, and the proposed} sequential procedure is \purple{also evaluated with its} correct allocation probability.}

\subsection{Application to Logistic Models}
\label{sec:simu}
Suppose $Y_k=1 (0)$ denotes a response variable with success
(failure) from a subject assigned to treatment $k$ for $k=1,
\ldots, K$. Let $\mu_k(\theta_k, \boldsymbol\xi) = E[Y_k=1| \boldsymbol\xi]$, and
\purple{$\theta_k  = (\alpha_k, \theta_k^*)$.} Assume that
\begin{align}
  \mbox{logit}(\mu_k(\theta_k, \boldsymbol\xi)) = \alpha_k +\theta_k^*\boldsymbol\xi, \/ k=1, \ldots, K.
\end{align}
Since the covariate vector can be redefined as $(1, \boldsymbol\xi)'$, without loss of generality,
we assume that $\alpha_k=0$, $k=1, \ldots, K$. Suppose there are $m_0$
initial samples for each treatment and assume that we are at the
$m$th stage with $m > Km_0$. Then the MLE $\hat\theta_{m,k}$
of $\theta_k$, for $k=1, \ldots, K$, is the one that
maximizes
\begin{align}
  L_k = \prod_{i=1}^m \mu_{i,k}^{X_{i,k}Y_{i,k}}
  (1-\mu_{i,k})^{X_{i,k}(1-Y_{i,k})},
\end{align}
where $\mu_{i, k}= \mu_k(\theta_k, \blue{\boldsymbol\xi_i})$. It follows  that the
conditional Fisher information matrix, for given $\boldsymbol\xi$, is
\[
    I_k(\theta_k|\boldsymbol\xi) =\mu_k(\theta_k, \boldsymbol\xi) (1-\mu_k(\theta_k, \boldsymbol\xi)) \boldsymbol{\xi\xi{'}} .
\]
Let $\hat I_{n,k} = n^{-1} \sum_{i=1}^n X_{i,k}
I_k({\hat{\theta}_{n,k}}|\boldsymbol\xi_i)$ be the estimate of $I_k$ for
all $k$. Then for a $K$ treatments problem, for example, the new design is
chosen such that the Fisher information matrix $\hat I_{n+1}$ is
maximized, if we assume $\eta=0$, where $\hat I_{n+1} =\hat I_{n}+\hat I^{n+1}$,
{
\begin{align}\label{information-matrix}
     &\hat I_{n} =\left(
          \begin{array}{ccc}
            \frac{1}{n} \sum_{i=1}^n X_{i,1} \hat\lambda_{i,1}{\boldsymbol\xi}_i{\boldsymbol\xi_i}' & 0 & 0 \\
            0 & \ddots & 0 \\
            0 & 0 & \frac{1}{n} \sum_{i=1}^n X_{i,K} \hat\lambda_{i,K}{\boldsymbol\xi}_i{\boldsymbol\xi_i}' \\
         \end{array}
        \right),\notag   \\
        &\hat I^{n+1} \equiv \left(
          \begin{array}{ccc}
            p_1\hat{\lambda}_{j,1}{\boldsymbol\xi}_j{\boldsymbol\xi_j}' & 0 & 0 \\
            0 & \ddots & 0 \\
            0 & 0 & p_K\hat{\lambda}_{j,K}{\boldsymbol\xi}_j{\boldsymbol\xi_j}' \\
           \end{array}
        \right),
\end{align}
}
and $\hat{\lambda}_{i,k}=\hat \mu_{i,k}(1-\hat \mu_{i,k})$ $\mbox{for } i=1, \ldots, n,$ $j=n+1$, and $k=1, \ldots, K$.

\subsubsection{Parameter Setup and Simulation Results}
Suppose that $K=2$; that is, we assume logistic models with binary responses,
\blue{two} treatments and one continuous covariate $\blue{\xi}$.  In the logistic models, we assume equal intercepts for both treatments
$(\alpha_{1},  \alpha_2) = (0.1, 0.1)$ and regression coefficients
$\blue{(\theta_1^*, \theta_2^*)} = (-1, 1)$. The covariate is generated from a mixed
normal distribution with means 2 $\& -2$ and equal variance 1 with
respective probability 0.5. Since the treatment effect is defined
as a function of differences of intercepts and regression
coefficients between the two treatments, we apply the stopping
rule for the contrasts of parameters, $\boldsymbol\gamma=H^{'} \boldsymbol\theta$, given
in Section \ref{s:subset}. Thus, the transpose of the contrast $H$
is defined as a matrix with its first row $(1, -1, 0, 0)$ and its
second row $(0, 0, 1, -1)$,
and the vector of parameters $\boldsymbol\theta$ is $\blue{{(\alpha_1 , \alpha_2 , \theta_1^* , \theta_2^* )}}^{'}$.

Precision $\delta$ is assumed $0.3$ and initial sample size for each treatment, $m_0$, is assumed as $5, 10,$ and $15$.
Several combinations of tuning parameters $T_0$ and $\eta$ are
assumed: $0.5, 1$ and $2$ for $T_0$ and $0, 0.1$ and $1$ for $\eta$.
Both fixed and varying tuning parameters, $T_0$ and $\eta$, are considered;
that is,
$T_0$ and $\eta$ are fixed until the study stops, or
vary whenever a new observation is added in a way that $T_0$ is proportional and $\eta$ is inversely proportional to the standard deviation of the treatment effect for a given covariate of a new observation.
Findings from simulation studies are as follows:



\begin{table}[b]
 \vspace*{-6pt}
 \centering
 \def\~{\hphantom{0}}
\caption{\em Mean (M) and standard deviation (SD) of stopping time ($\tau_{\delta_{\gamma}}$), coverage probability (CP) and correct allocation probability (CAP) of sequential $95\%$ confidence interval estimation with $\delta=0.3$.
$T_{0V}$ and $\eta_V$ indicate whether $T_0$ and $\eta$ vary or not}
\label{tab:fully-seq-subset}
\begin{scriptsize}
\begin{tabular}{lllll@{}rrrrr@{}rlllll@{}rrrrr}
\hline
     &      &        &\multicolumn{2}{c}{Variation}   && \multicolumn{2}{c}{$\tau_{\delta_{\gamma}}$} &    &      &      &        &&        &\multicolumn{2}{c}{Variation} && \multicolumn{2}{c}{$\tau_{\delta_{\gamma}}$} &    &      \\
\cline{4-5}\cline{7-8}  \cline{15-16}\cline{18-19}
$m0$ & $T_0$ & $\eta$ & $T_{0V}$ & $\eta_V$ && {M}                  & SD         & CP & CAP && $m0$ & $T_0$ & $\eta$ & $T_{0V}$ & $\eta_V$ && {M}                  & SD         & CP & CAP\\
\cline{1-5} \cline{7-10}               \cline{12-16} \cline{18-21}
 5 & 0.5 & 0.0 & N & N & & 53   &  9    & 0.95  & 0.48 & & 10 & 1.0 & 0.1 & Y & Y & & 60    & 11    & 0.96  & 0.87 \\
 5 & 0.5 & 0.0 & Y & N & & 58   & 24    & 0.90  & 0.46 & & 10 & 1.0 & 1.0 & N & N & & 60    & 11    & 0.98  & 0.92 \\
 5 & 0.5 & 0.1 & N & N & & 67   & 21    & 0.94  & 0.76 & & 10 & 1.0 & 1.0 & N & Y & & 63    & 14    & 0.96  & 0.92 \\
 5 & 0.5 & 0.1 & N & Y & & 67   & 23    & 0.93  & 0.74 & & 10 & 1.0 & 1.0 & Y & N & & 65    & 16    & 0.98  & 0.93 \\
 5 & 0.5 & 0.1 & Y & N & & 76   & 34    & 0.92  & 0.88 & & 10 & 1.0 & 1.0 & Y & Y & & 63    & 14    & 1.00  & 0.93 \\
 5 & 0.5 & 0.1 & Y & Y & & 66   & 16    & 0.97  & 0.87 & &    &     &     &   &   & &       &       &       &      \\
 5 & 0.5 & 1.0 & N & N & & 83   & 59    & 0.92  & 0.76 & & 10 & 2.0 & 0.0 & N & N & & 55    & 14    & 0.92  & 0.47 \\
 5 & 0.5 & 1.0 & N & Y & & 70   & 33    & 0.95  & 0.77 & & 10 & 2.0 & 0.0 & Y & N & & 52    & 14    & 0.97  & 0.46 \\
 5 & 0.5 & 1.0 & Y & N & & 81   & 20    & 0.96  & 0.91 & & 10 & 2.0 & 0.1 & N & N & & 59    & 13    & 0.96  & 0.83 \\
 5 & 0.5 & 1.0 & Y & Y & & 75   & 18    & 0.96  & 0.91 & & 10 & 2.0 & 0.1 & N & Y & & 57    & 12    & 0.99  & 0.83 \\
   &     &     &   &   & &      &       &       &      & & 10 & 2.0 & 0.1 & Y & N & & 54    & 12    & 0.96  & 0.76 \\
 5 & 1.0 & 0.0 & N & N & & 53   & 13    & 0.96  & 0.50 & & 10 & 2.0 & 0.1 & Y & Y & & 56    & 10    & 0.97  & 0.77 \\
 5 & 1.0 & 0.0 & Y & N & & 54   & 13    & 0.94  & 0.49 & & 10 & 2.0 & 1.0 & N & N & & 62    & 18    & 0.95  & 0.87 \\
 5 & 1.0 & 0.1 & N & N & & 69   & 17    & 0.92  & 0.87 & & 10 & 2.0 & 1.0 & N & Y & & 66    & 22    & 0.92  & 0.88 \\
 5 & 1.0 & 0.1 & N & Y & & 68   & 15    & 0.96  & 0.85 & & 10 & 2.0 & 1.0 & Y & N & & 58    & 12    & 1.00  & 0.84 \\
 5 & 1.0 & 0.1 & Y & N & & 66   & 15    & 0.96  & 0.84 & & 10 & 2.0 & 1.0 & Y & Y & & 57    & 14    & 0.96  & 0.83 \\
 5 & 1.0 & 0.1 & Y & Y & & 66   & 16    & 0.86  & 0.84 & &    &     &     &   &   & &       &       &       &      \\
 5 & 1.0 & 1.0 & N & N & & 75   & 19    & 0.96  & 0.91 & & 15 & 0.5 & 0.0 & N & N & & 52    & 10    & 0.98  & 0.50 \\
 5 & 1.0 & 1.0 & N & Y & & 77   & 18    & 0.98  & 0.91 & & 15 & 0.5 & 0.0 & Y & N & & 51    &  9    & 0.96  & 0.45 \\
 5 & 1.0 & 1.0 & Y & N & & 72   & 13    & 0.96  & 0.90 & & 15 & 0.5 & 0.1 & N & N & & 52    &  9    & 0.98  & 0.72 \\
 5 & 1.0 & 1.0 & Y & Y & & 73   & 19    & 0.99  & 0.90 & & 15 & 0.5 & 0.1 & N & Y & & 53    & 10    & 0.98  & 0.74 \\
   &     &     &   &   & &      &       &       &      & & 15 & 0.5 & 0.1 & Y & N & & 58    & 12    & 0.98  & 0.87 \\
 5 & 2.0 & 0.0 & N & N & & 54   & 11    & 0.90  & 0.48 & & 15 & 0.5 & 0.1 & Y & Y & & 55    & 14    & 0.98  & 0.90 \\
 5 & 2.0 & 0.0 & Y & N & & 53   & 10    & 0.94  & 0.49 & & 15 & 0.5 & 1.0 & N & N & & 54    & 10    & 1.00  & 0.75 \\
 5 & 2.0 & 0.1 & N & N & & 60   & 14    & 0.99  & 0.79 & & 15 & 0.5 & 1.0 & N & Y & & 54    & 18    & 0.98  & 0.76 \\
 5 & 2.0 & 0.1 & N & Y & & 61   & 14    & 0.96  & 0.79 & & 15 & 0.5 & 1.0 & Y & N & & 57    & 20    & 0.94  & 0.90 \\
 5 & 2.0 & 0.1 & Y & N & & 54   & 10    & 0.98  & 0.73 & & 15 & 0.5 & 1.0 & Y & Y & & 56    & 18    & 0.96  & 0.89 \\
 5 & 2.0 & 0.1 & Y & Y & & 57   & 13    & 0.94  & 0.72 & &    &     &     &   &   & &       &       &       &      \\
 5 & 2.0 & 1.0 & N & N & & 73   & 20    & 0.91  & 0.86 & & 15 & 1.0 & 0.0 & N & N & & 52    &  9    & 0.96  & 0.50 \\
 5 & 2.0 & 1.0 & N & Y & & 71   & 18    & 0.97  & 0.88 & & 15 & 1.0 & 0.0 & Y & N & & 52    & 10    & 0.97  & 0.43 \\
 5 & 2.0 & 1.0 & Y & N & & 59   & 14    & 0.96  & 0.80 & & 15 & 1.0 & 0.1 & N & N & & 56    & 11    & 0.98  & 0.91 \\
 5 & 2.0 & 1.0 & Y & Y & & 57   & 11    & 0.98  & 0.78 & & 15 & 1.0 & 0.1 & N & Y & & 55    & 13    & 0.98  & 0.90 \\
   &     &     &   &   & &      &       &       &      & & 15 & 1.0 & 0.1 & Y & N & & 55    & 11    & 0.98  & 0.92 \\
10 & 0.5 & 0.0 & N & N & & 52   &  8    & 0.92  & 0.48 & & 15 & 1.0 & 0.1 & Y & Y & & 53    & 12    & 0.96  & 0.91 \\
10 & 0.5 & 0.0 & Y & N & & 52   & 10    & 0.92  & 0.46 & & 15 & 1.0 & 1.0 & N & N & & 54    & 11    & 0.96  & 0.91 \\
10 & 0.5 & 0.1 & N & N & & 58   & 12    & 0.96  & 0.74 & & 15 & 1.0 & 1.0 & N & Y & & 58    & 18    & 0.96  & 0.93 \\
10 & 0.5 & 0.1 & N & Y & & 59   & 16    & 0.98  & 0.75 & & 15 & 1.0 & 1.0 & Y & N & & 57    & 12    & 0.94  & 0.95 \\
10 & 0.5 & 0.1 & Y & N & & 63   & 14    & 0.96  & 0.90 & & 15 & 1.0 & 1.0 & Y & Y & & 56    & 10    & 0.98  & 0.95 \\
10 & 0.5 & 0.1 & Y & Y & & 61   & 13    & 1.00  & 0.91 & &    &     &     &   &   & &       &       &       &      \\
10 & 0.5 & 1.0 & N & N & & 57   & 13    & 0.99  & 0.78 & & 15 & 2.0 & 0.0 & N & N & & 52    &  9    & 0.97  & 0.49 \\
10 & 0.5 & 1.0 & N & Y & & 57   & 14    & 0.97  & 0.74 & & 15 & 2.0 & 0.0 & Y & N & & 52    &  9    & 0.96  & 0.44 \\
10 & 0.5 & 1.0 & Y & N & & 64   & 16    & 0.93  & 0.92 & & 15 & 2.0 & 0.1 & N & N & & 55    & 15    & 0.96  & 0.88 \\
10 & 0.5 & 1.0 & Y & Y & & 61   & 12    & 0.98  & 0.92 & & 15 & 2.0 & 0.1 & N & Y & & 54    & 11    & 1.00  & 0.84 \\
   &     &     &   &   & &      &       &       &      & & 15 & 2.0 & 0.1 & Y & N & & 53    & 10    & 0.97  & 0.83 \\
10 & 1.0 & 0.0 & N & N & & 54   & 11    & 0.92  & 0.46 & & 15 & 2.0 & 0.1 & Y & Y & & 50    &  8    & 0.96  & 0.79 \\
10 & 1.0 & 0.0 & Y & N & & 52   &  9    & 0.96  & 0.47 & & 15 & 2.0 & 1.0 & N & N & & 59    & 15    & 0.98  & 0.90 \\
10 & 1.0 & 0.1 & N & N & & 63   & 16    & 0.98  & 0.90 & & 15 & 2.0 & 1.0 & N & Y & & 54    & 12    & 0.95  & 0.88 \\
10 & 1.0 & 0.1 & N & Y & & 62   & 17    & 0.94  & 0.87 & & 15 & 2.0 & 1.0 & Y & N & & 55    & 12    & 0.98  & 0.86 \\
10 & 1.0 & 0.1 & Y & N & & 63   & 14    & 0.96  & 0.89 & & 15 & 2.0 & 1.0 & Y & Y & & 53    & 13    & 0.98  & 0.85 \\ \hline   \end{tabular} \vskip18pt
\end{scriptsize}
\end{table}

As $\eta$ gets larger, stopping time gets larger but its increase
is reduced as initial sample size gets larger. Varying $\eta$
does not give results that are significantly different  from fixed
$\eta$ unless $T_0$ varies as well.
Stopping time is very unstable when initial sample size $m_0$ is
small, such as $5$, due to unstable regression coefficient
estimates at the beginning stage if $T_0$ is 0.5 or 2.
%
%
As initial sample size gets larger, stopping time gets earlier and its variation gets smaller.
The coverage probabilities of treatment differences are reasonably
close to the nominal level 0.95 and become closer to 0.95 as the initial sample
size $m_0$ gets larger.
\blue{Based on these findings, it is recommended that, in order to obtain earlier stable stopping time with a given precision satisfied, \red{the initial sample size should not be too small.}}

\blue{When $\eta=0$,  correct treatment allocation probabilities are about 0.5, since it is equivalent to randomized allocation as there is no ethical consideration in the utility function.}
As $\eta$ gets larger, correct treatment allocation gets better
with similar performance for positive $\eta$. This confirms that
$\eta$ plays a role as a tuning parameter for ethical
consideration and a, small, nonzero $\eta$ is sufficient for correct
allocation. \blue{Large correct allocation probabilities for positive $\eta$, in Table \ref{tab:fully-seq-subset}, illustrate that our sequential procedure under the CARA designs successfully implements the idea of CARA designs, with more allocation to better treatment, for the non-sequential counterpart.}

For positive $\eta$, correct allocation is high and close to 0.9
when $T_0 =1$ or when initial $T_0$=0.5 with varying $T_0$. However,
it is lower when $T_0 = 0.5$ with fixed $T_0$ compared to varying
$T_0$ or when $T_0 = 2$ with varying $T_0$ compared to fixed
$T_0$. If $T_0$ varies depending on treatment effect variation,
$T_0$ becomes larger than the initial $T_0$. Thus, varying small
$T_0$ gives better allocation due to the reasonable tuning size of
$T_0$, however, varying large $T_0$ gives worse allocation due to a
too liberal tuning of $T_0$. This emphasizes the importance of
selecting a reasonably sized $T_0$.
%
%

\renewcommand{\thesection}{\arabic{section}.}
\setcounter{section}{4}
\section{Discussion}\label{sec: D}
\renewcommand{\thesection}{\arabic{section}}

In this paper, we propose a sequential estimation scheme for the CARA design in clinical trials. \purple{In this sequential estimation procedure,} allocation function and design depend not only on previously collected information and sequential estimates of treatment effects, but also on the covariate information of individual subjects. \purple{The proposed sequential estimation is based on the martingale estimating equation, which differs from some classical sequential methods that rely on independent observations.}  \purple{The stopping rule used here depends on the observed Fisher information, which guarantees the precision of the estimates of treatment effects, and is novel in the CARA design based clinical trials. The procedure discussed here is rather general and can be applied to other generalized linear models.}  We demonstrate our method \purple{using} some logistic regression models \purple{under} a two-treatment case.
The theorems derived in this work are \purple{for general allocation rules, which require only mild conditions on the allocation function}. It will be possible to explore more if a specific allocation rule is available.

As shown in Figure 1, it is very difficult to allocate the most suitable treatment for subjects in the vicinity of the intersection of lines of two treatment effects. \purple{This is especially the case, when} the difference in slopes of treatments is \purple{small}. Thus, instead of a strictly concave function as we have used in our numerical study, some concave function with a plateau may be considered. According to our experience based on the numerical studies\purple{, the large changes adopted in $T_0$ during the sequential procedure may lower} the correct allocation \blue{probability}. \purple{ Hence, from a practical viewpoint, a reasonably sized $T_0$ should be chosen in the utility function, considering all factors of a clinical trial, such as the distributions of covariates, the intersection point of the two treatment models and among others}.   In other words, if we have some prior information on the targeted sub-populations, then it may help to decide $T_0$.  This leads to possible future research, where Bayesian statistical tools might play an important role.
\section*{Acknowledgement}
This work was partially supported by the Korea Research Foundation (KRF) grant funded by the Government of Korea (MEST) (R01-2009-0076473), and by Taiwan National Science Council (NSC 99-2118-M-001-001).


\renewcommand{\thesection}{\Alph{subsection}}
\setcounter{section}{0}
\setcounter{theorem}{0}
\vspace{\fill}\pagebreak

\section*{Appendix A}
To apply  sequential sampling to CARA designs, we need to extend
the results of Anscombe's theorem to daptive design. From the
proof of Anscombe's theorem (see \shortciteNP{woodroof82}, page
11), the i.i.d. assumption is not necessary; in fact, it only requires
the sequence of partial sum to satisfy the u.c.i.p. condition.
This is sufficient for applying  Anscombe's theorem.
The lemma below shows that the sequence of the partial sum of martingale
differences also satisfies the u.c.i.p. condition.  The arguments
below are similar to those of \shortciteN{woodroof82}, example
1.8.
\begin{lemma}\label{lemma:ucip-martingale} Let $X_1, X_2,
\ldots$ be a sequence of martingale differences with respect to a
sequence of increasing $\sigma$-field ${\cal F}_i$ for $i=0,1,
\ldots $; that is, $E[X_{i} | {\cal F}_{i-1}]=0$ for all $i \geq
1$. Suppose that there is a constant $M$ such that $E[\| X_{i}
\|^2 | {\cal F}_{i-1}] < M < \infty$ for all $i$. Then
$Y_n=S_n^*=\sum_{i=1}^n X_i/\sqrt{n}$ satisfies the u.c.i.p.
condition.
\end{lemma}
\noindent{\bf Proof of Lemma \ref{lemma:ucip-martingale}}\\
For all $k, n \geq 1$, $|S_{n+k}^* - S_n^*| \leq \sqrt n |S_{n+k} - S_n |+ \left[ 1 + \sqrt{\frac{n}{n+k}}\right] | S_n^*|$, where $S_n = \sum_{i=1}^n X_i $. If $\epsilon, \delta > 0$ and $k \leq n \delta$, then the second term on the right hand side is bounded by $C(\delta) |S_n^*|$, where $C(\delta) = 1 - (1+\delta)^{-1/2}$ and
  \[
    P\left( C(\delta) |S_n^*| > \frac \epsilon 2 \right) \leq P
    \left( |S_n^*| > \frac \epsilon {2 C(\delta)}\right) \rightarrow
    0 \hbox{ as } \delta \rightarrow 0,
  \]
since $|S_n^*|$ is stochastically bounded.
Because $X_i$'s are martingale differences, instead of Komogorov's inequality, we apply the H\'{a}jek-R\'{e}ney inequality (see \shortciteNP{ChowTeicher88}, Theorem 8 (iii), page 247). Then it is shown that
  \[
    P\left(\max_{k \leq n \delta}|S_{n+k} - S_n| \geq \frac{\epsilon\sqrt
    n}{2} \right) \leq \left(\frac 4 {n \epsilon ^2}\right) n\delta M = \frac{4\delta
    M}{\epsilon^2},
  \]
which is independent of $n$ and goes to zero as $\delta \rightarrow 0$. Therefore, $S_n^*$, $n \geq 1$, satisfies the u.c.i.p. condition.

\subsection{Last time for generalized linear models}\label{subsect:last time}

We can apply the last time method for martingale differences as that in \shortciteN{chang99} in our proof of asymptotic efficiency.

Let $\sigma^2_i \equiv
\hbox{Var}(Y_i| \boldsymbol\xi_i) = \sigma^2 \nu(\mu_k)$. Then for fixed
$\rho
>0$ and for each $k$, let's define a last time variable
\[
    L_{k, \rho} = \sup\{n \geq 1: (\theta - \theta_k)' \ell_n(\theta) > 0 \, \exists \, \theta \,\in \partial {{\mathbf
    \Theta}}_{k, \rho}
    \},
\]
where $\ell_{n, k}(\theta) = \sum_{i=1}^n
g(\xi_i'\theta_k)\xi_{i}(Y_{i,k} - \mu_k(\theta_k, \xi_i)$ and
$g(t) = \dot \mu_k/\nu(\mu_k)$, provided that the derivative of
$\mu_k$ exists. Then it follows from \shortciteN{chang99},
\[
    n > L_{k, \rho} \Rightarrow \hat\theta_k \hbox{ exists and }
    \hat \theta_k \in \partial {\mathbf \Theta}_{k, \rho}.
\]
Moreover, he proved that under some regularity conditions of
covariate $\boldsymbol\xi$'s, $EL_{k, \rho} < \infty$ for all $k$. This
implies that if we define the last time $L_\rho = \max\{L_{1, \rho},
\ldots, L_{K, \rho}\}$, then $n > L_\rho$ implies that $\hat\theta
\in {\mathbf \Theta}_\rho \subset {\mathbf \Theta}$, where
$\Theta_\rho = \prod_{k=1}^K {\mathbf \Theta}_{k, \rho}$.

Note that \shortciteN{chang99} defined  last times for generalized linear models, and it is clear
that for each $k \in \{1, \ldots, K \}$, Equation (\ref{model-1}) is a special case of
\shortciteN{chang99}. In \shortciteN{ZFCC07}, they assume the estimate of $\boldsymbol\theta$ exists in a
compact set when sample size $n$ is sufficiently large.  By the last time defined above, since we can choose sufficiently small $\rho$ such that for sufficiently large $n$,
$\hat {\boldsymbol\theta}$ will fall into a compact neighborhood of $\boldsymbol\theta$. (Hence, the assumption of \shortciteN{ZFCC07} can be relaxed. See \shortciteN{chang99} for further details).

  Although the treatment allocation for each subject is affected by
  previous observed responses, it is clear that the estimate of
  $\theta_k$, $k=1, \ldots, K$,
  is still calculated separately for a given sample under the general CARA design.
  Thus the estimation procedure of $\theta_k$'s for all different $k$'s can be treated as estimating $K$
  adaptive regression models, separately.  That is, for given observations, the estimation
  of $\theta_k$, for each $k$, does not depend on estimates of
  other $\theta_l$, $l \not = k$. That is, if for  $k=1, \ldots, K$, let
  \[
    U_{m,k}= \hbox{ collection of observations } \{Y_{j,k}, \blue{\boldsymbol\xi}_{j} \hbox{ with } X_{j,k}=1: j=1,\ldots, m\},
  \]
  then the estimate of $\theta_k$, say $\hat\theta_k$, is calculated based on observations in $U_{m,k}$ only.
  Thus, the property of $\hat\theta_k$ is the same as the MLE of a stochastic regression model.
  The sequential estimate under the adaptive design has been studied by some authors. For example,
  \shortciteN{Lai-n-Wei-1982} studied its properties under a linear regression setup with a general adaptive design assumption, while \shortciteN{ChenHuYing99}, and \shortciteN{chang01}
  discussed estimation under a generalized linear model setup.  Their results are applied in the proof of Theorem \ref{thm:seq-main}. (In these three papers, they only assume that the design is adaptive, but no particular design scheme is assumed. Hence, their methods are rather general and can be applied to our case under some specific allocation rules.)\\

\noindent{\bf Proof of Theorem \ref{thm:seq-main}}\\
It is proved in \shortciteN{ZFCC07} that $\hat \V$ is a strongly consistent estimate of $\V$. This
implies that \red{$\Lambda_{\max}(\hat \V)$ and $\Lambda_{\min}(\hat \V)$} are also  strongly consistent estimates of \red{$\Lambda_{\max}(\V)$ and $\Lambda_{\min}(\V)$, respectively.}  Thus, if
{$\sup_m \|  \red{\boldsymbol\xi_m} \| < \infty$}, then by \shortciteN{chow-robbins65}, Lemma 1, it is shown that
$P(\tau_\delta < \infty) =1$ and $\lim_{d \rightarrow 0} \tau_\delta / n_{opt} =1$
with probability one and thus the proof of (i) is completed.

The highlight of the proof of (ii) is the asymptotic normality
under the random sample size. This property can usually be
obtained  by applying Anscombe's Theorem, which relies on the u.c.i.p.
property (see \shortciteNP{woodroof82}). However, under the
adaptive design, some modification is required.  Thus, here we
apply its modification, which is stated as Lemma
\ref{lemma:ucip-martingale}.

The asymptotic normality of $\boldsymbol {\hat\theta}$ under the adaptive design has been established by
\shortciteN{ZFCC07} (see also \shortciteNP{Lai-n-Wei-1982}, \shortciteNP{chang99} and
\shortciteNP{ChenHuYing99}). Following from the results of (i), to prove (ii), it suffices to prove
that the sequence of normalized random sums {$\{ \sqrt n(\boldsymbol{\hat \theta}_n - \boldsymbol\theta), n \geq 1 \}$}
satisfies the u.c.i.p condition (see \shortciteNP{woodroof82} for its definition).
From Equation (2.4) of \shortciteN{ZFCC07}, we have, with probability one, $\hat\theta_{n, k} - \theta_{k} = n^{-1}\sum_{m=1}^nX_{m,k} h_k(Y_{m,k}, \boldsymbol\xi_m)(1+o(1)) + o(n^{-1/2})$, where function $h_k$ satisfies $E[h_k(Y_k, \boldsymbol\xi)|\boldsymbol\xi]=0$.

{It follows from Zhang et al. (2007) that we have  $$\sqrt n (\hat\theta_{n, k} - \theta_{k})=n^{-1/2}\sum_{m=1}^nX_{m,k} h_k(Y_{m,k}, \blue{\boldsymbol\xi}_m)+o(1) n^{-1/2} \sum_{m=1}^nX_{m,k} h_k(Y_{m,k}, \blue{\boldsymbol\xi}_m)+o(1)$$ almost surely. } It is clear that from the definition of u.c.i.p., the property of convergence with probability one will imply the property of u.c.i.p. Moreover, it follows from Lemma 1.4 of \shortciteN{woodroof82}, if both $U_n$ and $W_n$ are u.c.i.p., then $U_n+W_n$ is also u.c.i.p.
{By applying Lemma \ref{lemma:ucip-martingale}, we have that $\{n^{-1/2}\sum_{m=1}^nX_{m,k} h_k(Y_{m,k}, \blue{\boldsymbol\xi}_m):  n \geq 1\}$ is u.c.i.p., which together with Lemma 1.4 of \shortciteN{woodroof82} implies that $\{\sqrt n (\hat\theta_{n, k} - \theta_{k}): n \geq 1\} $ satisfies the u.c.i.p. condition.}
Hence, applying Anscombe's theorem (Theorem 1.4 of \shortciteN{woodroof82}; see also Theorem 4.5.3 of \shortciteN{govin04}), the asymptotic normality of $\hat\theta_{n, k}$ remains for each $k$, and it completes the proof of (ii).

It follows from (i), that to prove {(iii), it suffices} to prove that $\{\delta \T_\delta: \delta \in (0,1)\}$ is uniformly integrable. As discussed in Section \ref{subsect:last time},
\[
 n > L_{\rho} \Rightarrow \boldsymbol{\hat\theta}_n \in {\boldsymbol \Theta}.
\]
Since ${\mathbf \Theta}$ is compact, this implies that for $n > L_{\rho}$, $\Lambda_{\max}(\hat V_k)
\leq \sup_{\theta \in {\mathbf \Theta}}\Lambda_{\max}(V_k(\theta))\leq C_{\Lambda_{\max}}$ for some
$C_{\Lambda_{\max}} > 0$.
Let $\tilde V_k=\hbox{diag}\{O,\ldots, \hat V_k, \ldots, O \}$ for $k=1, \ldots, K$, where $O$
denotes the $p\times p$ matrix of $0$'s. Then $\hat \V=\sum_{k=1}^K \tilde V_k$. Thus,
$\Lambda_{\max}(\hat \V) \leq \sum_{k=1}^K \Lambda_{\max}(\tilde V_k) \leq K C_{\Lambda_{\max}}$.
Hence, for $n > L_\rho$, the stopping time $\T_\delta$ is bounded. Moreover, by applying
the last time lemma for martingale differences in \shortciteN{chang99},
we have $E[L_\rho]< \infty$. This implies that $\{\delta \T_\delta: \delta
\in (0,1)\}$ is uniformly integrable and the proof of (iii) is completed.

The proofs of (iv) and (v) follow directly from {Theorem 2.1, Equation (2.6) and Theorem 2.2,
Equation (2.8)} of \shortciteN{ZFCC07} and the strong consistency of $\tau_\delta$, so they are
omitted here.
{To prove (vi), we only need to show that $\{n^{-1/2} ({\mathbf N}_n - {n} \boldsymbol\nu) : n= 1,2,
\cdots. \}$ is u.c.i.p.} {From (A.6) of \shortciteN{ZFCC07}, we have,} with probability
one,
$${\mathbf N}_n - n\boldsymbol\nu = {\mathbf M}_n + (1+o(1))\sum_{i=1}^n
\sum_{k=1}^K \frac{T_{i,k}}{m}(\partial {\boldsymbol\nu/\partial {\mathbf \theta}_k})' + o(n^{1/2}),
$$
{where ${\mathbf M}_n=(\Delta M_{n,1},\ldots, \Delta M_{n,K})$ and ${\mathbf T}_n = (\Delta T_{n,1}, \ldots, \Delta T_{n, K})$} are multi-dimensional martingale sequences with bounded martingale differences; that is, $\Delta M_{n,k} \leq 1$ and $\|\Delta {\mathbf T}_{n,k}\| < \infty$, where  $\Delta$  denotes the operand of a sequence $\{z_n\}$; that is, $\Delta z_n = z_n - z_{n-1}$.  (Here only the moment condition of martingale differences is required for our purpose. Thus, other properties of $\mathbf M_n$ and $\mathbf T_n$ are omitted. See \shortciteN{ZFCC07} for further details.) Therefore, with probability one,
\[
    n^{-1/2} ( \N_n - n\boldsymbol\nu) = n^{-1/2} \mathbf M_n {+} (1+o(1)) n^{-1/2} \sum_{i=1}^n
    \left[\sum_{k=1}^K \frac{T_{i,k}}{m}(\partial {\boldsymbol\nu/\partial {\mathbf \theta}_k})'\right] + o(1).
\]
{Similarly,} by applying Lemma \ref{lemma:ucip-martingale} {again and arguments similar to} \shortciteN{woodroof82}, Example 1.8, we have $\{n^{-1/2} ({\N}_n -{n \boldsymbol\nu}) : n= 1,2, \cdots. \}$ is u.c.i.p. This completes the proof of Theorem \ref{thm:seq-main} (vi). \\

\noindent{\bf Proof of Theorem \ref{thm:subset}}\\
 By the definition of $\boldsymbol\gamma = H' \boldsymbol\theta$ and $rank(H) = h \leq p$, it easy to see that $\hat{\boldsymbol\gamma}$ is a strongly consistent
  estimate of $\boldsymbol\gamma$ and is asymptotically normally distributed with covariance matrix $V_{\gamma}$.  Moreover, it is clear that
  $\{\sqrt n (\boldsymbol{\hat\gamma}_n - \boldsymbol\gamma) : n= 1,2, \cdots. \}$ is u.c.i.p., since $H$ is a non-random matrix.
  Thus, the proofs of Theorem \ref{thm:subset} (i) and (ii) follow from the same arguments as in the proofs of Theorem \ref{thm:seq-main} (i) and (ii).
  To prove (iii), we first note that by simple matrix algebra, we have
  \[
    \Lambda_{\max}(\hat \V_\gamma) =  \Lambda_{\max}(H'\hat \V H) \leq \Lambda_{\max}(H'H) \cdot \Lambda_{\max}(\hat \V).
  \]
  Since $H$ is a pre-fixed non-random matrix, $\Lambda_{\max}(H'H)$($=\lambda_H$, say) is a constant. Now, let
  \[
    {\tilde\T}_{\delta_\gamma}
           = \inf\{ n \geq n_0 : n \geq \lambda_H \frac{C^2_{\alpha, \gamma}
           \Lambda_{\max}({ \hat \V_\gamma} )}{\delta_\gamma^2}\}.
  \]
  Then by definition, we have $ \T_{\delta_\gamma} \leq {\tilde\T}_{\delta_\gamma} $ almost everywhere. Moreover, again it can be shown by the same arguments above that $\{ d^2 {\tilde\T}_{\delta_\gamma}: d \in (0,1)\}$ is uniformly integrable. This implies that
  $\{ d^2 \T_{\delta_\gamma}: d \in (0,1)\}$ is uniformly integrable and thus the proof of (iii) of Theorem \ref{thm:subset} is completed.

\bibliographystyle{chicago}
\bibliography{scara}

\newpage
\begin{figure}\centering\label{fig:SCARA} 
  \includegraphics[width=12cm, height=10cm]{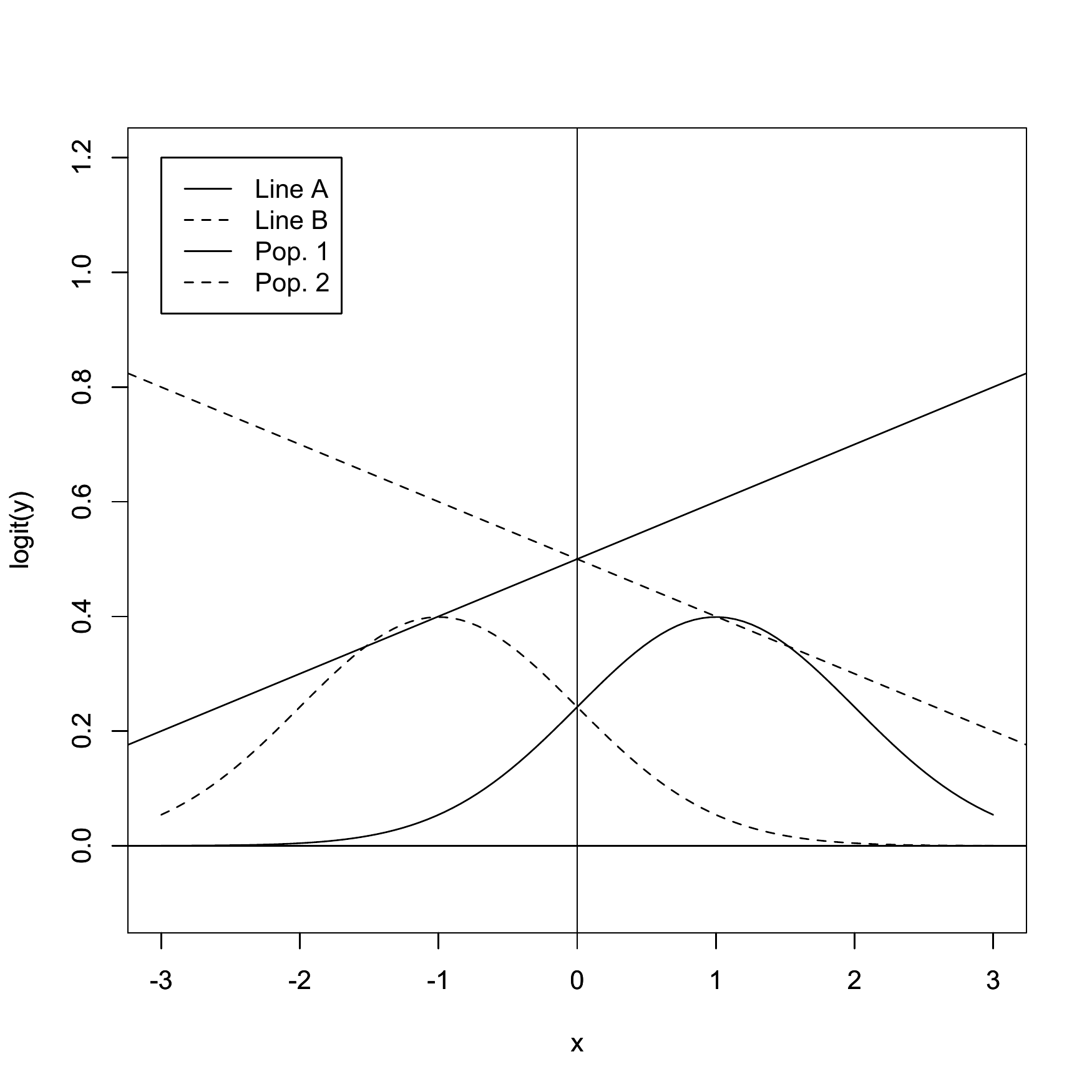}
  \caption{{Sequential CARA design with two treatment slopes {and two covariate populations}.}}
\end{figure}

\end{document}